\documentclass[conference]{IEEEtran}
\IEEEoverridecommandlockouts
\usepackage{cite}
\usepackage[english]{babel}

\usepackage[T2A,LGR,T1]{fontenc}
\usepackage[latin9]{inputenc}
\usepackage{array}
\usepackage{float}
\usepackage{times}
\usepackage[letterpaper, left=0.62in, right=0.62in, top=0.74in, bottom=1.06in]{geometry}

\usepackage{textcomp}
\usepackage{multirow}
\usepackage{stackrel}
\usepackage{graphicx}
\usepackage{booktabs}
\usepackage{setspace}
\usepackage[export]{adjustbox}
\usepackage{algorithm}
\usepackage{caption}
\usepackage{subcaption}
\usepackage{algpseudocode}
\usepackage{enumerate}
\usepackage{xcolor}
\usepackage{graphicx}
\usepackage{amssymb}
\usepackage{color}
\usepackage{amsmath,amssymb,amsfonts}
\def\BibTeX{{\rm B\kern-.05em{\sc i\kern-.025em b}\kern-.08em
    T\kern-.1667em\lower.7ex\hbox{E}\kern-.125emX}}

\begin{document}

\title{The Interplay of AI-and-RAN: Dynamic Resource Allocation for Converged 6G Platform}


\author{
  Syed Danial Ali Shah\textsuperscript{$\star$} \quad Zeinab Nezami\textsuperscript{$\star$} \quad Maryam Hafeez\textsuperscript{$\star$} \quad Syed Ali Raza Zaidi\textsuperscript{$\star$}\\ 
  \textsuperscript{$\star$}School of Electrical and Electronic Engineering, University of Leeds, Leeds, UK\\
  \textsuperscript{$\star$}Corresponding author: Syed Ali Raza Zaidi (s.a.zaidi@leeds.ac.uk)
}

\maketitle
\footnotetext{\textit{This paper has been accepted for presentation at the IEEE INFOCOM 2025 Workshop and will appear in the IEEE INFOCOM 2025 proceedings.}}


\maketitle

\begin{abstract}
The concept of AI-RAN as specified by the AI-RAN alliance is geared to explore a converged 6G platform that can support management, orchestration, and deployment of both AI and RAN workloads. This concept is central to the development of a 6G architecture that aims to exploit the accelerated compute capabilities for supporting both real-time signal processing and offloading of Generative AI (GenAI) workloads.  However, both the architectural framework required to support this vision and the dynamic resource allocation strategy are still in their infancy. The O-RAN architecture intrinsically allows cloud-native disaggregated implementation. Consequently, we explore a framework that can allow orchestration of AI-and-RAN workloads by expanding the Near Real-Time RAN Intelligent Controller (NRT-RIC) within O-RAN.  The framework incorporates a monitoring xApp that tracks RAN KPIs and exposes radio analytics to the proposed E2E orchestrator via a recently introduced Y1 interface. The orchestrator implements a Soft Actor-Critic (SAC) reinforcement learning algorithm to dynamically allocate critical computing resources, e.g., Multi-Instance GPUs (MIGs), between latency-sensitive RAN network functions and computationally intensive AI workloads on shared RAN infrastructure.  The proposed framework provides insight on how the traditional RAN architecture can be evolved to inherently support emerging GenAI workloads. Our framework prioritizes the real-time requirements of RAN workloads while maintaining efficient resource sharing for AI applications. The simulation results demonstrate the benefits of the proposed framework, as it meets nearly 99\% of the requests for RAN workload while effectively supporting AI workloads and achieving 100\% utilization of the RAN infrastructure resources in a dynamic environment.
\end{abstract}


\section{Introduction}
The rapid expansion of mobile communications and the growing demand for network capacity is driving innovations in NextG wireless network architecture. To improve resource utilisation and harness maximum value from the scarce spectral resource, increased virtualisation, disaggregation, and densification have become core ingredients of network design. While there are several alternative architectural frameworks, one possible architectural choice that has democratised the research for Next-G wireless networks is Open Radio Access Network (O-RAN) architecture. O-RAN has transformed traditional cellular networks by enabling open interfaces and multi-vendor interoperability, paving the way for efficient resource utilization and dynamic network optimization \cite{danial11}. However, these advancements also introduce new challenges as networks evolve to support increasingly diverse workloads. In particular, with the growing deployment of AI \& ML applications alongside traditional RAN workloads, there is a pressing need for intelligent resource management systems that can effectively share computing infrastructure between these diverse workload types \cite{danial2,danial3}. Current RAN infrastructures often maintain dedicated resources for network functions, leading to under-utilisation during off-peak periods while simultaneously struggling to accommodate the growing demands of AI workloads.

AI and RAN, one of the three key domains envisioned by the AI-RAN Alliance, aims to support the coexistence of AI and RAN workloads on shared RAN infrastructure \cite{danial3}. This involves deploying AI applications on RAN infrastructure, leveraging shared accelerated compute (e.g. GPU) and memory resources to run intensive AI tasks while simultaneously supporting RAN operations \cite{danial3}.  The coexistence of RAN and AI workloads on shared infrastructure presents unique challenges due to their distinct characteristics, dynamic nature, and lifecycle management requirements. Cellular networks are inherently dynamic where RAN workloads continuously vary, making real-time tracking of resource demands essential for optimal resource allocation, in particular for the co-existence scenarios. In contrast, AI workloads are computationally intensive and exhibit variable resource demands depending on their specific use cases and deployment contexts. Traditional static resource allocation approaches and architectural frameworks are inadequate for addressing the nuanced requirements of AI and RAN coexistence on shared infrastructure.

\begin{figure*}[hbt!]
	\centering
	\includegraphics[width=0.70\textwidth]{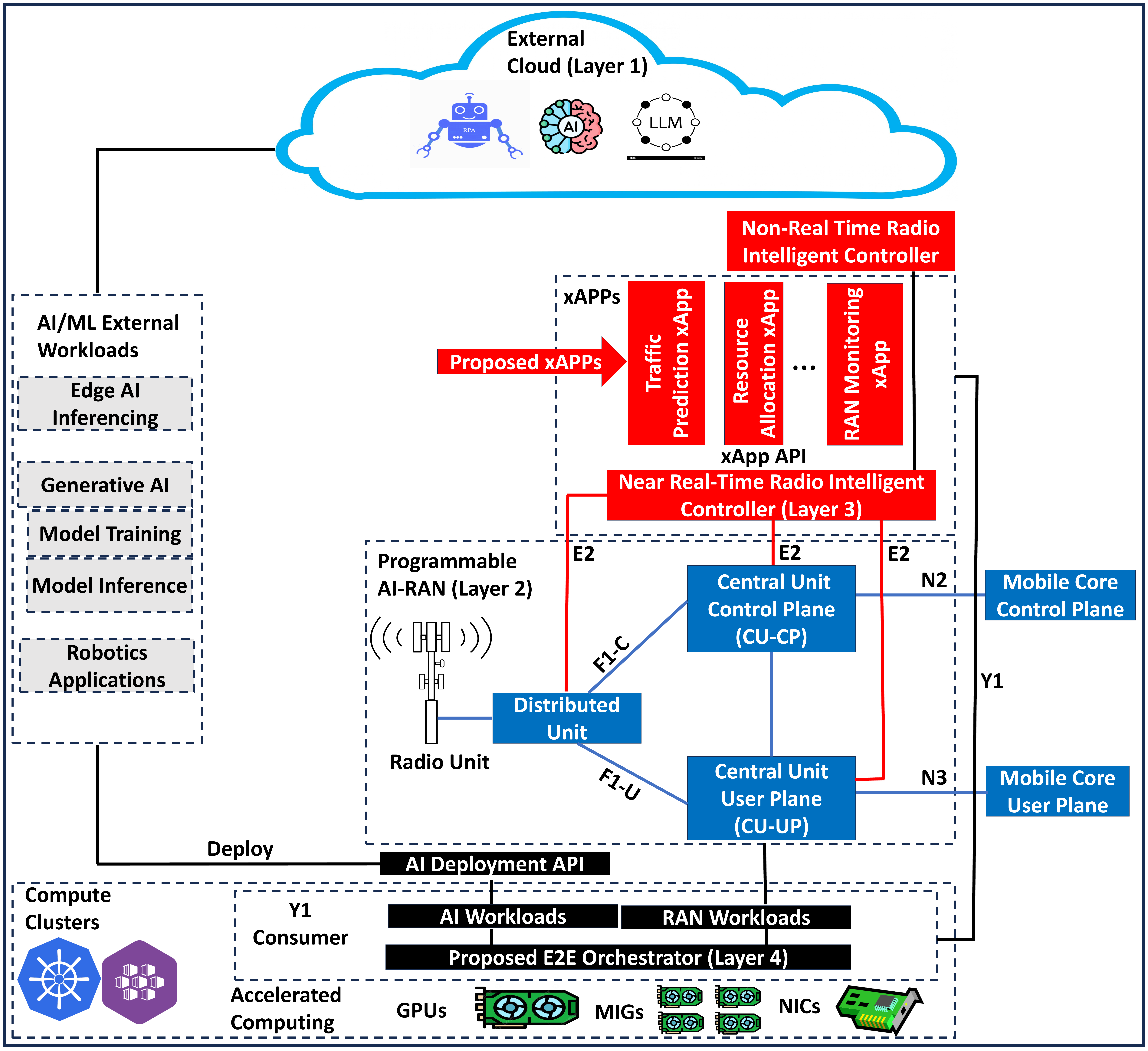}
\captionsetup{justification=raggedright,singlelinecheck=false,labelsep=period,font=small}
	\caption{Converged AI-and-ORAN architecture} 
	\label{fed}%
\end{figure*}

Some recent advances and AI-based solutions have shown promise in dynamic resource management scenarios. However, existing solutions focus either on RAN optimization or AI workload scheduling independently and are insufficient for AI and RAN coexistence scenarios. The authors in \cite{danial4} presented a comprehensive power consumption model for O-RAN configurations to understand the energy consumption patterns in O-RAN architecture. In \cite{danial5}, authors presented Radio Intelligent Controller (RIC) modules integrating meta-learning for real-time network data collection and dynamic management of RAN resources. An O-RAN-compatible adversarial learning-based resource allocation scheme is proposed in \cite{danial6} to enhance the energy efficiency of virtualised base stations and O-RAN components. Some other relevant works introduce advanced O-RAN testbeds and frameworks, e.g., Open AI Cellular (OAIC) framework for designing and testing AI-based RAN Management Algorithms \cite{danial7}, and X5G, a private 5G O-RAN Testbed with NVIDIA ARC and OpenAirInterface \cite{danial8,danial9}.

However, none of the above state-of-the-art works and frameworks address the architectural framework or optimization requirements necessary to support the coexistence of AI and RAN on shared infrastructure. Furthermore, the dynamic nature of cellular networks, the real-time demands of RAN network functions, and the varying computational needs of AI workloads necessitate a more sophisticated approach capable of dynamically adjusting and adapting resource allocation policies based on the changing requirements.

This paper presents a novel architectural framework based on O-RAN specifications to address the requirements for dynamic resource allocation and the coexistence of RAN and AI workloads on shared RAN infrastructure. The key working principles of the proposed framework and its contributions are summarized as follows:

\begin{itemize}
    \item \textbf{Converged AI-and-ORAN Architectural Framework:} To explore the convergence of both compute-and-communication, we propose a Converged AI-and-ORAN Architectural (CAORA) framework based on O-RAN specifications, incorporating custom-built xApps within the Near-Real-Time RAN Intelligent Controller (NRT-RIC). This framework advances the evolving AI and RAN domain by enabling the coexistence of AI and RAN workloads on shared RAN infrastructure. The xApp continuously tracks RAN KPIs and exposes the radio analytics information with the proposed End-to-End (E2E) orchestrator through the recently introduced Y1 interface, enabling dynamic data collection and network monitoring for intelligent resource management decisions.
    
    \item \textbf{Intelligent E2E Orchestrator and Cross-Layer Resource Management Framework:} We introduced a system that establishes communication between xApps integrated within the NRT-RIC and the external E2E orchestrator via the Y1 interface. This framework enables real-time adaptation of resource allocation policies based on evolving network dynamics. The proposed E2E orchestrator employs a Soft Actor-Critic (SAC) reinforcement learning algorithm, acting as a Y1 consumer to communicate with the NRT-RIC. It processes real-time metrics from the monitoring xAPP, maintains historical network patterns, and implements dynamic resource allocation policies, thereby enabling the coexistence of RAN and AI workloads on shared RAN infrastructure.
    
    \item \textbf{Implementation and Performance Analysis:} We demonstrated the effectiveness of our proposed framework through simulation results, which highlight the maximised utility of the shared RAN infrastructure. The results validate the system's capability to maintain RAN performance requirements in dynamic scenarios while concurrently supporting the coexistence of AI workloads on shared infrastructure resources.
\end{itemize}

The remainder of this paper is organized as follows: Section II presents the proposed converged architecture. Section III details the system model and problem formulation. Section IV discusses the simulation environment and performance evaluation. Finally, Section V concludes the paper.

\section{Converged AI-and-ORAN Architectural Framework}
We present the CAORA framework to enable the coexistence of RAN and AI workloads on shared RAN infrastructures, as shown in Fig.~\ref{fed}. CAORA implements a hierarchical control structure comprising four key layers: The external cloud, programmable RAN components, the controller, and resource management layers. The external cloud layer (layer 1) hosts AI/ML workloads, e.g., edge AI inferencing, generative AI, and robotics applications, that interface with the RAN infrastructure, i.e., compute clusters, through the AI Deployment API. The programmable RAN layer (layer 2) includes disaggregated RAN components based on the O-RAN architecture, e.g., Distributed Units (O-DU), O-RAN Radio Units (O-RU), and O-RAN Central Units (O-CU).  The controller layer (layer 3) consists of the NRT-RIC, which integrates with the proposed xApps for traffic prediction, resource allocation, and network monitoring, facilitating rapid resource allocation decisions via the E2 interface. The proposed xApps, e.g., monitoring xApp, are also responsible for consistently tracking and monitoring the network load information to determine the RAN workload requirements and communicating it with the proposed E2E orchestrator. 

The E2E Orchestrator implements a SAC reinforcement learning algorithm interfacing with the monitoring xApp, deployed within the NRT-RIC through the Y1 interface, as shown in Fig.~\ref{fed}. The Y1 interface connects the NRT-RIC with an authorized Y1 consumer for exposure to radio analytics information. The monitoring xApp continuously collects critical network metrics, e.g., bandwidth utilization, system latency, and network load, and feeds this data into the SAC agent, i.e., the Y1 consumer. The SAC agent uses this information to implement the priority-based and dynamic allocation of shared computing resources, where RAN workloads receive precedence during peak periods while AI workloads utilise available resources during off-peak times, supporting the coexistence of the RAN and AI workloads on shared RAN infrastructure, as illustrated in Fig.~\ref{fed}. The proposed SAC agent supports the coexistence of RAN and AI workloads on shared infrastructure through three key mechanisms: predictive analysis utilizing historical patterns for resource planning, real-time adaptation adjusting allocations based on current conditions as informed by the monitoring xApp, and priority management ensuring RAN performance while maximizing AI workload support. 

 The resource management layer (layer 4) implements the proposed E2E orchestrator that manages the RAN infrastructure resources, i.e., compute clusters containing GPUs, Multi-Instance GPUs (MIGs) based on NVIDIA's MIG technology \cite{danial13}, and Network Interface Cards (NICs). The proposed architectural framework transforms traditional RAN infrastructure into a multi-purpose system that efficiently supports both network operations and AI workloads.
 
\section{System Model}
\subsection{AI and RAN Workloads: Representation and Task Categories}
Let $K$ denote the set of all tasks in the system, including RAN and AI workloads. Each task $k \in K$ is represented as:
\begin{equation}
k = \{ \tau, \boldsymbol{r}, p \},
\end{equation}

\noindent where $\tau$ is the type of task, e.g., RAN or AI workload task, $\boldsymbol{r}=[r_i,\forall i=1,2,...,N]$ represents the resource requirements with $r_i$ being requirement for resource type $i$, and $p$ is the priority level, i.e., $p\in[0,1]$. The priority levels are defined to allow the system to account for different types of tasks, e.g., within RAN workloads, some tasks may require higher priorities, with real-time tasks like signal processing taking precedence over non-real-time tasks like traffic analysis or reporting. Notice that the resource demand dynamically varies with time \(t\), i.e., $\boldsymbol{r}(t)$ with $t\in \mathbb{R}$. Without loss of generality, we denote the $\boldsymbol{r}(t)\in k$ as $\boldsymbol{r}(k,t)$ and $p\in k$ as $p(k)$.

The resource demand for an AI/RAN task \( k \) at time \( t \) is defined as the sum of its individual resource requirements across all resource types:

\begin{equation}
\Delta(k, t) = \lVert \boldsymbol{r}(k,t)\rVert_1 = \sum_{i=1}^{n} r_i(k, t),
\end{equation}

\noindent where \( r_i(k, t) \) represents the resource requirement of task \( k \) for resource type \( i \) at time \( t \). Consequently, the total resource demand across all tasks in the system at time \( t \), denoted as \( \Delta_{\text{total}}(t) \), is given by:

\begin{equation}
\Delta_{\text{total}}(t) = \sum_{k \in K} \Delta(k, t).
\end{equation}

\subsection{System Dynamics}
The system dynamics define the total workload evolution by balancing task generation, task completion, and resource contention over time. The total workload \( W(t) \) is expressed as:



\begin{equation}
\small
W(t) = \int_0^t \left( \sum_{k \in K} P(k, t) \Delta(k, t) 
- \sum_{k \in K} P_c(k, t) \Delta(k, t) C(k, t) \right) dt
\end{equation}

\noindent where \( P(k, t) \) represents the probability of task generation for task \( k \) at time \( t \), \( P_c(k, t) \) denotes the probability of task completion for task \( k \) at time \( t \), defined as \( P_c(k, t) = \alpha \cdot p(k) \), with \( \alpha \) being the system efficiency coefficient and \( p(k) \) the task priority level, and \( C(k, t) \) represents resource contention for task \( k \), defined as $C(k, t) = \frac{\Delta_{\text{total}}(t)}{\mathcal{R}_{\text{max}}}$. Here, \( \mathcal{R}_{\text{max}} \) represents the maximum available resources. 

\begin{algorithm}
\caption{SAC-Based Dynamic Resource Allocation for RAN and AI Workloads}
\label{alg:sac_allocation}
\begin{algorithmic}[1]
\State \textbf{Input:} $d_{RAN}(t)$, $d_{AI}(t)$, $R_{max}$
\State \textbf{Output:} Optimal $r_{RAN}(t)$, $r_{AI}(t)$

\State \textbf{Initialize:} NRT-RIC, xApp, E2E Orchestrator, SAC agent ($\pi$, $Q_1$, $Q_2$), Replay Buffer $D$

\For{each time step $t$}
    \State \textbf{1. Collect KPIs:} $KPIs \gets$ xApp (latency, throughput, network load)
    \State \textbf{2. Update State:} $s_t = \{d_{RAN}, d_{AI}, r_{RAN}, r_{AI}\}$
    \State \textbf{3. SAC Policy Decision:} $a_t = \pi(s_t)$
    \State \textbf{4. Resource Allocation:}
    \State $r_{RAN}(t) \gets r_{RAN}(t-1) + \Delta r_{RAN}(t)$
    \State $r_{AI}(t) \gets r_{AI}(t-1) + \Delta r_{AI}(t)$
    \State s.t. constraint $r_{RAN}(t) + r_{AI}(t) \leq R_{max}$
    \State \textbf{5. Calculate Reward:} $R_t'$
    \State \textbf{6. Update Replay Buffer:} $D \leftarrow (s_t, a_t, R_t', s_{t+1})$
    \State \textbf{7. SAC Training:}
    \State Sample $(s, a, R, s')$ from $D$
    \State Update $Q_i$: $Q_i(s, a) = R + \gamma \mathbb{E}[Q_i(s', a')]$
    \State Update $\pi$: $\nabla_\theta J(\pi) \approx \mathbb{E}[\nabla_\theta \pi(s) Q(s, a)]$
\EndFor

\State \textbf{Return:} Optimal $r_{RAN}(t)$, $r_{AI}(t)$
\end{algorithmic}
\end{algorithm}

\subsection{Problem Formulation}
We model the resource allocation problem as a dynamic system in which the maximum available resources \( \mathcal{R}_{\text{max}} \) are dynamically distributed between RAN and AI workloads. Let \( r_{\text{RAN}}(t) \) and \( r_{\text{AI}}(t) \) denote the resource allocations at time \( t \) for RAN and AI workloads, respectively, such that:

\begin{equation} \label{eq:resource_constraint}
    r_{\text{RAN}}(t) + r_{\text{AI}}(t) \leq \mathcal{R}_{\text{max}},
\end{equation}

\noindent where \( \mathcal{R}_{\text{max}} = 1 \) represents the normalized maximum resources available and is defined as \( \mathcal{R}_{\text{max}} = R_{\text{base}} \cdot f(u) \). The scaling function \( f(u) \) is applied to adjust the available resource pool based on the number of active users \( u \) in the network. This parameter is consistently updated and tracked by the proposed monitoring xApp and fed into the SAC agent to make dynamic resource allocation decisions based on the varying resource demands.

The demand for RAN and AI workloads at time \( t \), denoted as \( d_{\text{RAN}}(t) \) and \( d_{\text{AI}}(t) \), respectively, is given by \( d_{\text{RAN}}(t) = \Delta_{\text{total}}^{\text{RAN}}(t) \) and \( d_{\text{AI}}(t) = \Delta_{\text{total}}^{\text{AI}}(t) \), and represents the total resource requirement for RAN and AI tasks at time \( t \). The completion rates for RAN and AI workloads at time \( t \), denoted as \( C_{\text{RAN}}(t) \) and \( C_{\text{AI}}(t) \), represent the proportion of the total resource demand for RAN and AI tasks that has been fulfilled. These completion rates are defined as:

\begin{equation} \label{eq:completion_rates_combined}
    C_x(t) = \min \left( p_x(k) \cdot r_x(t), d_x(t) \right), \quad x \in \{\text{RAN}, \text{AI}\}.
\end{equation}

\noindent where \( p_{\text{x}}(k) \) denotes the priority levels of RAN or AI tasks.

\subsection{Dynamic Resource Adaptation}
The proposed system implements the SAC algorithm, based on reinforcement learning, to address dynamic fluctuations in resource demands by optimally and dynamically adjusting the resource allocations \( r_{\text{RAN}}(t) \) and \( r_{\text{AI}}(t) \). The model-specific components are highlighted as follows:

\subsubsection{States}
The state \( s_t \) at time \( t \) encapsulates the system's demand and resource utilization. The state is defined as follows:
\begin{equation}
    s_t = \{d_{\text{RAN}}(t), d_{\text{AI}}(t), r_{\text{RAN}}(t-1), r_{\text{AI}}(t-1)\}
\end{equation}

\noindent where \( d_{\text{RAN}}(t) \) and \( d_{\text{AI}}(t) \) represent the current demands for RAN and AI workloads, respectively, while \( r_{\text{RAN}}(t-1) \) and \( r_{\text{AI}}(t-1) \) denote the resource allocations from the previous time step. The inclusion of past allocations in the system model captures the system's temporal dynamics, thereby facilitating more informed decision-making.

\subsubsection{Actions}
The action \( a_t \) represents the adjustment in resource allocation for RAN and AI workloads, and is defined as follows:
\begin{equation}
    a_t = \{\Delta r_{\text{RAN}}(t), \Delta r_{\text{AI}}(t)\}
\end{equation}

\noindent where \( \Delta r_{\text{RAN}}(t) \) and \( \Delta r_{\text{AI}}(t) \) represent the increments or decrements in resource allocation for RAN and AI workloads, respectively, as determined by the SAC policy.

\subsubsection{Reward Function}
We formulated the reward function to achieve optimal resource allocation and maximize the utility of shared RAN infrastructure resources. The reward function includes time-varying weights $w_{\text{RAN}}(t)$ and $w_{\text{AI}}(t)$ to balance the completion of RAN and AI workloads and prioritize one over another in case of certain events, e.g., peak demand for RAN. These weights are dynamically tuned by the proposed SAC agent, enabling real-time adjustments for optimal resource allocation policies. The base reward function \( R_t \) is defined as:

\begin{equation} \label{eq:utility_function}
    R_t = w_{\text{RAN}}(t) \frac{C_{\text{RAN}}(t)}{d_{\text{RAN}}(t)} + w_{\text{AI}}(t) \frac{C_{\text{AI}}(t)}{d_{\text{AI}}(t)}.
\end{equation}

\noindent where the weights are normalized such that \( w_{\text{RAN}}(t) + w_{\text{AI}}(t) = 1 \). For efficient utilization of system resources, we extended the base reward function by introducing a penalty term that accounts for underutilised resources. The updated reward function is defined as:

\begin{equation}
    R_t' = R_t - \lambda \cdot \left( 1 - \frac{R_{\text{allocated}}(t)}{R_{\text{max}}} \right)
\end{equation}

\noindent Where $\lambda$ represents the penalty coefficient, and $R_{allocated}(t)=r_{\text{RAN}}(t) + r_{\text{AI}}(t)$ denotes the total resources allocated at time $t$ for both AI and RAN workloads. The SAC algorithm aims to optimize cumulative utility over a time horizon \( T \), which is expressed as:

\begin{equation}
    \mathcal{U} = \mathbb{E}\left[\sum_{t=0}^{T} \gamma^t  R_t'\right]
\end{equation}

\noindent where \( \gamma \) represents the discount factor.

The pseudocode for the proposed CAORA framework is shown in Algorithm 1.

\begin{table}[!t]
\caption{Simulation Parameters}
\label{tab:params}
\centering
\begin{tabular}{|l|l|}
\hline
\textbf{Parameter} & \textbf{Value} \\
\hline
Training Episodes & 1000 \\
Time Steps per Episode & 100 \\
Off-peak RAN/AI Demand Range & [2, 5] MIGs \\
Peak RAN Demand (Congested) Range & [6, 7] MIGs \\
Compute Cluster Resources ($R_{max}$) & 7 MIGs \\
Actor/Critic Learning Rate & 3e-4 \\
Discount Factor ($\gamma$) & 0.99 \\
Temperature Parameter ($\alpha$) & 0.2 \\
Replay Buffer Capacity & 100,000 \\
Neural Network Hidden Layer Size & 128 \\
Training Batch Size & 64 \\
Actor/Critic Architecture & 3-layer MLP \\
\hline
\end{tabular}
\end{table}

\section{Performance Evaluation}
\subsection{Experimental Setup and Implementation}
The experimental evaluation of the proposed CAORA framework was performed using Python with the O-RAN 7.2x split \cite{danial11}. We simulated realistic O-RAN infrastructure dynamics with shared computing resources between the RAN workloads and the AI workloads. The proposed monitoring xApp is integrated with the NRT-RIC as a Python module that consistently tracks the network updates, e.g., resource demands \( d_{\text{RAN}}(t) \), from the micro-cell served by gNBs (i.e., 5G NR), and feeds this information to the proposed E2E orchestrator. The orchestrator dynamically allocates resources to RAN workloads based on real-time network load updates while efficiently distributing any remaining resources to AI workloads within the shared O-RAN infrastructure. The proposed E2E orchestrator is based on the SAC agent's neural architecture and consists of actor-network and dual critic networks to enhance policy learning stability and state-value estimation accuracy. 

In this paper, resources are modelled as MIGs based on Nvidia MIG technology \cite{danial13}, where each GPU instance possesses dedicated resources that can be allocated to specific workloads, such as AI or RAN tasks. NVIDIA's MIG technology enables partitioning a single GPU into isolated instances with dedicated memory, cache, and compute cores, ensuring efficient resource utilization \cite{danial13}. The resource configuration for the simulation environment is modelled using the MIG 1g.5gb profile of the NVIDIA A100-SXM4-40GB GPU, which supports up to 7 MIG instances, each with 1/8 of the total GPU memory allocated \cite{danial13}. For this evaluation, we assumed that each task, i.e., AI or RAN, requires an equal allocation of resources for completion, specifically 1 MIG per task at a time.  In future work, we aim to enhance the proposed model by incorporating variable resource allocations. 

\begin{figure}[]
	\centering
	\includegraphics[width=0.47\textwidth]{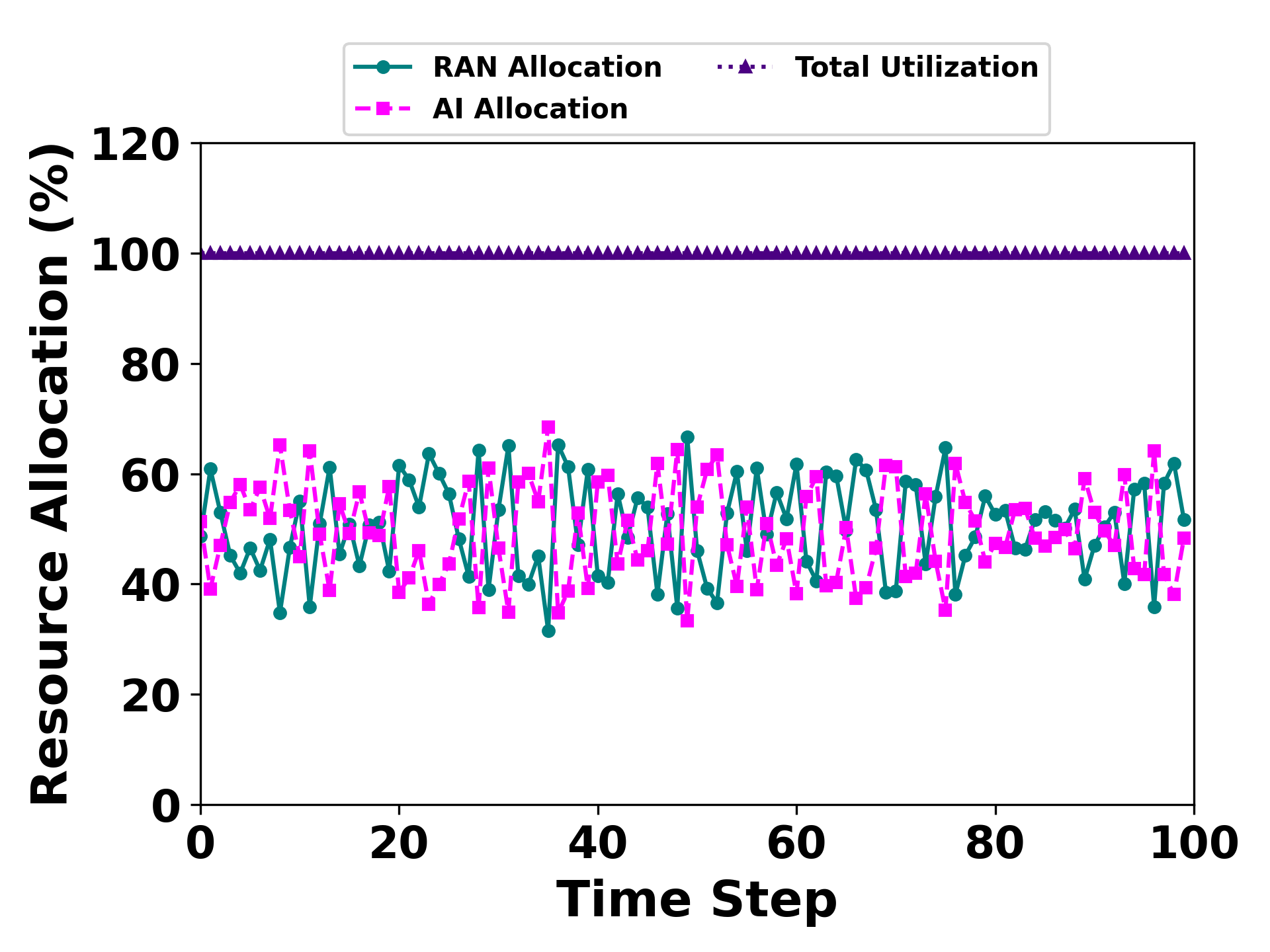}
\captionsetup{justification=raggedright,singlelinecheck=false,labelsep=period,font=small}
	\caption{Dynamic resource allocation between AI and RAN workloads in off-peak scenarios} 
	\label{workload_balance}%
\end{figure}

\begin{figure}[]
	\centering
	\includegraphics[width=0.47\textwidth]{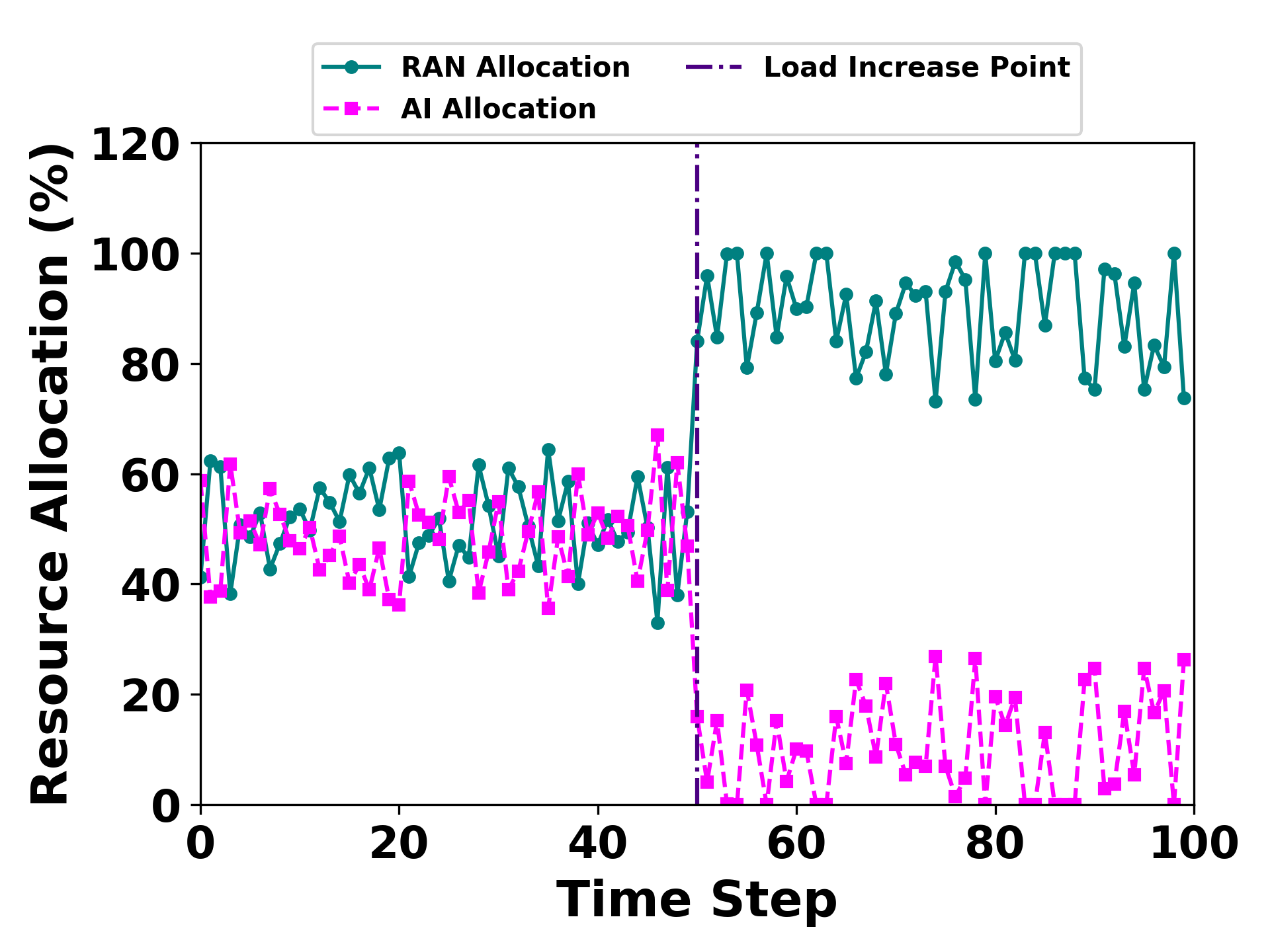}
\captionsetup{justification=raggedright,singlelinecheck=false,labelsep=period,font=small}
	\caption{Dynamic resource adaptation during peak RAN demand (congested) scenarios} 
	\label{utilization}%
\end{figure}

 We conducted a preliminary evaluation of the proposed model to assess its performance in effectively utilizing shared RAN infrastructure resources and supporting the coexistence of RAN and AI workloads on shared infrastructures. We simulated varying resource demands for AI and RAN workloads, $d_{\text{AI}}(t)$ and $d_{\text{RAN}}(t)$, respectively. This included MIG demands per time slot for off-peak and peak (congested) scenarios, which correspond to the fluctuating number of users in the network at any given time. In this research, we assumed that the fluctuating number of users is represented by MIG demands, as tracked by the xApp. In future work, we aim to test the proposed model on real datasets and heterogeneous measurements collected from different cellular base stations in the city, such as the dataset from Barcelona City presented in \cite{danial14}. Table \ref{tab:params} details the simulation parameters used throughout our evaluation.

\subsection{Simulation Results}
\subsubsection{Workload Balance Dynamics and Resource Utilization Analysis}
Fig. \ref{workload_balance} illustrates the temporal evolution of resource distribution between RAN and AI workloads tested on the offline-trained model. As varying workload demands for both RAN and AI were simulated during off-peak times, the proposed SAC agent dynamically allocated resources to each workload according to their respective demands. Since this simulation represents an off-peak scenario, the monitoring xApp integrated within the NRT-RIC updates the SAC-based E2E orchestrator to effectively balance resource allocation between RAN and AI workloads, ensuring efficient utilization of shared RAN infrastructure resources, as shown in Fig. \ref{workload_balance}. 


Furthermore, with the proposed CAORA framework, the system achieves 100\% utilization of computing resources, compared to the scenario where only RAN workloads are supported on the RAN infrastructure, as shown in Fig.~\ref{workload_balance}. This demonstrates the framework's ability to maintain high resource utilization while ensuring RAN service quality, making it particularly well-suited for practical O-RAN deployments where infrastructure efficiency is critical.

\subsubsection{Adaptive Resource Allocation and Task Completion}
Fig. \ref{utilization} illustrates the system's response to dynamic workload variations, particularly during critical load transition periods. The vertical demarcation at timestep 50 indicates the artificial injection of increased RAN demand, simulating a peak (congested) RAN demand scenario to evaluate the system's adaptive capabilities. The proposed monitoring xApp integrated within the NRT-RIC promptly predicts the increased resource demand for RAN network functions and updates the E2E orchestrator via the Y1 interface. The SAC agent, recognizing the increased RAN workload demands, dynamically adjusts the resource allocation and reallocates resources to RAN workloads while reducing or eliminating those allocated to AI workloads, as shown in Fig. \ref{utilization}. This shows the effectiveness of our priority-weighted reward function, as defined in the proposed SAC system model, in preserving critical RAN performance during high-demand periods.

\begin{figure}[]
	\centering
	\includegraphics[width=0.45\textwidth]{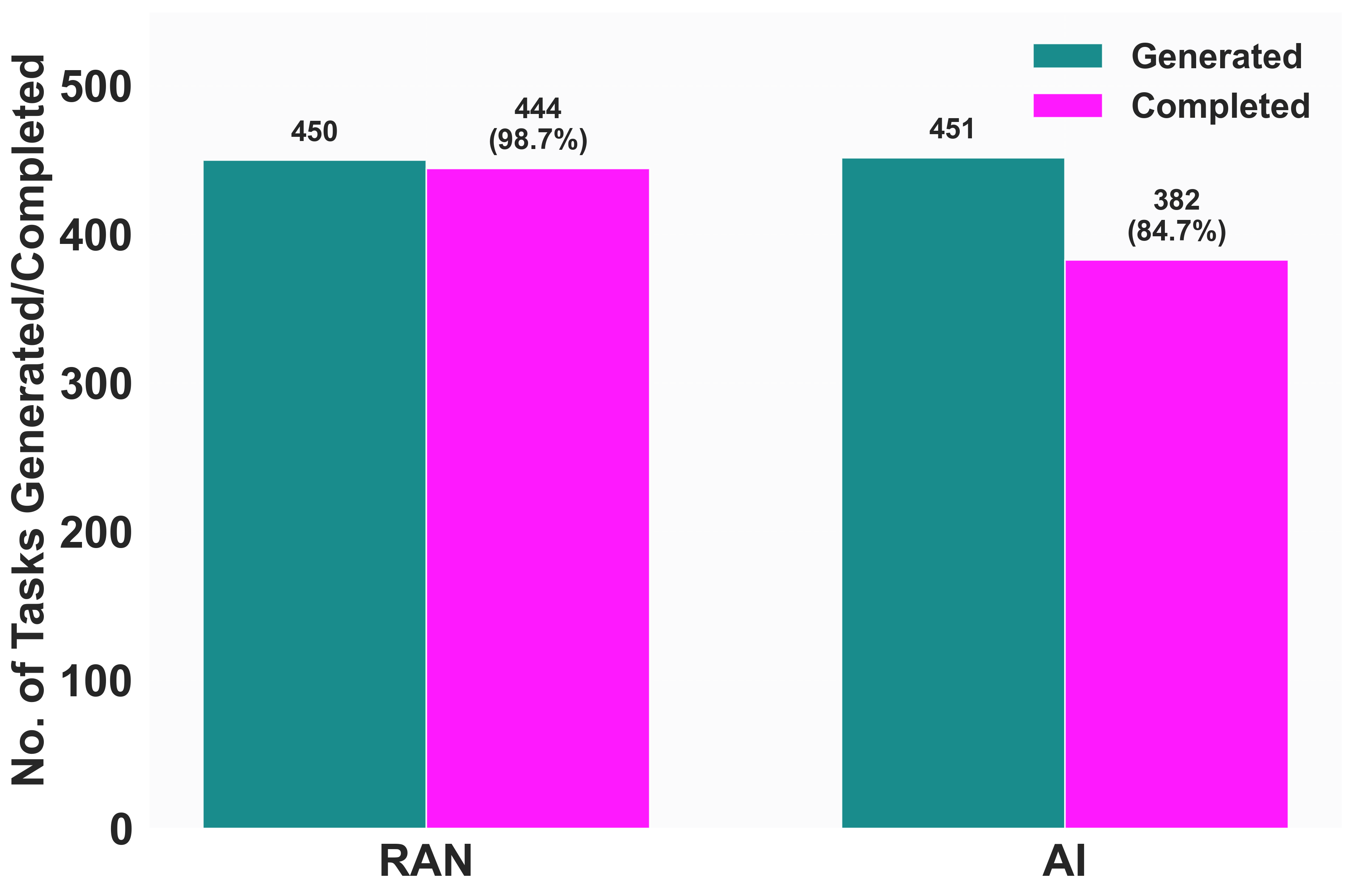}
\captionsetup{justification=raggedright,singlelinecheck=false,labelsep=period,font=small}
	\caption{Average task completion ratio (\%) per episode} 
	\label{taskcompletion}%
\end{figure}

Fig. \ref{taskcompletion} highlights the performance of the proposed system in terms of successfully completing the requested RAN and AI workloads/tasks during an off-peak scenario. The slightly lower task completion ratio for AI workloads reflects the dynamic resource allocation policy of the proposed system that emphasizes prioritizing RAN performance requirements, ensuring around 99\% completion ratio for RAN workloads while supporting the coexistence of both workloads.

\section{Conclusion}  
In this paper, we proposed a novel CAORA framework built on the O-RAN architecture to support the coexistence of RAN and AI workloads on shared infrastructure. We integrated a custom-built monitoring xApp within the NRT-RIC to monitor real-time KPIs and network updates and communicate with the proposed E2E orchestrator. The orchestrator implements the SAC reinforcement learning algorithm to dynamically allocate computing resources based on fluctuating workload demands for RAN and AI workloads. The proposed framework provides insights into transforming traditional RAN infrastructure by enabling multi-purpose utilization of shared resources, supporting the seamless coexistence of RAN and AI workloads, and optimizing resource efficiency without compromising RAN performance. The monitoring xApp provides continuous network insights, ensuring dynamic RAN requirements are met while surplus resources are allocated to AI workloads for better infrastructure capacity utilisation.


\section*{Acknowledgement}
This research was funded by EP/X040518/1 EPSRC CHEDDAR.


\bibliographystyle{IEEEtran}
\bibliography{references}

\end{document}